\documentclass[prd,aps,superscriptaddress,10pt,nobibnotes,notitlepage,twocolumn,longbibliography,nofootinbib,preprintnumbers]{revtex4-2}

\usepackage[normalem]{ulem}
\usepackage{sidecap}
\usepackage[latin1]{inputenc}
\usepackage{graphicx}
\usepackage{float}
\usepackage{latexsym}
\usepackage{graphicx}
\usepackage{amssymb}
\usepackage{amsmath,mathtools}
\usepackage{amsfonts}
\usepackage{tikz} 
\usepackage{ragged2e}
\usepackage{stackengine}
\usepackage{scalerel}

\usepackage[colorlinks=true,citecolor=blue,hyperfootnotes=false]{hyperref}

\usepackage{dcolumn}
\usepackage{textcomp}
\usepackage{xfrac}
\usepackage{slashed}
\usepackage{multirow}

\def\be{\begin{equation}}
\def\ee{\end{equation}}
\def\figs/B{B}
\def\bea{\begin{eqnarray}}
\def\eea{\end{eqnarray}}
\def\bg{\begin{eqnarray}}
\def\nd{\end{eqnarray}}

\def\beq{\begin{equation}}
\def\eeq{\end{equation}}

\renewcommand{\d}{{\rm d}}

\usepackage{accents}

\usepackage{graphicx}
\usepackage{dcolumn}
\usepackage{bm}
\usepackage{hyperref}
\usetikzlibrary{decorations.pathmorphing}
\tikzset{snake it/.style={decorate, decoration=snake}}
\newcommand{\sint}[3]{\hspace{- #3 em} \underset{#1}{\overset{#2}{\int}}\hspace{- #3 em}}

\usepackage{cleveref}

\usepackage{marginnote}

\definecolor{oiOrange}{RGB}{230,159,0}
\definecolor{oiBlue}{RGB}{0,114,178}
\definecolor{oiGreen}{RGB}{0,158,115}

\begin{document}

\preprint{MIT-CTP/5951}\preprint{IFT-UAM/CSIC-25-146}

\title{Path integral predictions for pre-asymptotic false vacuum decay}

\author{Joshua Lin}
 \email{joshua.lin@anl.gov}
\affiliation{%
Department of Physics, Massachusetts Institute of Technology, Cambridge, MA 02139, USA }
\affiliation{Physics Division, Argonne National Laboratory, Lemont, IL 60439, USA
}%

\author{Bruno Scheihing-Hitschfeld}
 \email{bscheihi@kitp.ucsb.edu}
\affiliation{%
Kavli Institute for Theoretical Physics, University of California, Santa Barbara, California 93106, USA
}%

\author{Thomas Steingasser}
 \email{thomas.steingasser@uam.es}
\affiliation{%
Department of Physics, Massachusetts Institute of Technology, Cambridge, MA 02139, USA }
\affiliation{Black Hole Initiative at Harvard University, 20 Garden Street, Cambridge, MA 02138, USA
}%
\affiliation{Departamento de Fisica Teorica, Universidad Autonoma de Madrid, and IFT-UAM/CSIC, Cantoblanco, 28049, Madrid, Spain
}%

\date{\today}

\begin{abstract}
When tunneling occurs out of generic initial states, a significant fraction of probability is lost at early times during which the dynamics is governed by excited resonance states. However, first-principles analyses based on path integrals have only captured the leading asymptotic behavior during which the tunneling rate is dominated by the false vacuum contribution. In this work, we discuss the behavior in the pre-asymptotic regime from a first-principles path integral perspective. We demonstrate how the relevant expressions can be evaluated systematically through semi-classical methods in the recently developed steadyon picture. This approach allows one to trace the role of the relevant physical scales, making transparent the underlying assumptions and approximations and offering a clear path to establishing a systematically improvable framework to evaluate tunneling rates non-perturbatively.
\end{abstract}

\maketitle

{\it \bf Introduction.} 
Quantum tunneling is one of the most universal features of quantum systems, governing phenomena that may take place at every length or energy scale, from nuclear decay rates to out-of-equilibrium phenomena in the early Universe. A common feature of many tunneling phenomena is that the systems of interest can rarely be found in an exact, idealized false vacuum state. And yet, modern first-principles path-integral based investigations of tunneling have only analyzed tunneling out of the latter state. While this is sufficient to describe the long-term asymptotics of the relevant probabilities, it fails to capture the dominant mechanism of probability loss at intermediate times. In this regime, the tunneling rate is dominated by contributions from excited states. 

In this Letter, we show how this behavior can be captured through a combination of the steadyon picture for the evaluation of real-time path integrals~\cite{Steingasser:2024ikl,Steingasser:2023gde} with the modern, first-principles based ``direct approach''~\cite{Andreassen:2016cvx,Andreassen:2016cff}, extending the advantages of this approach to the analysis of the early-time transient dynamics. While the specific details of the transient dynamics are highly dependent on the initial state, numerical investigations suggest that a significant fraction of the overall probability flux is lost during this transient period. Moreover, many applications rely on tunneling to occur during these early times. 

\begin{figure*}
    \centering
    \includegraphics[width=0.75\linewidth]{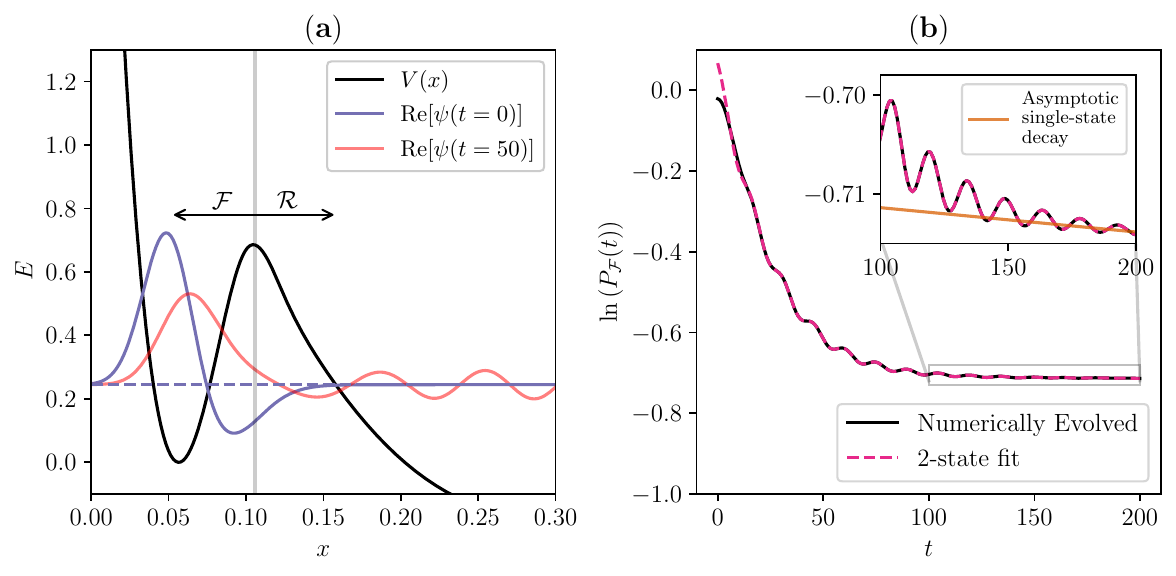}
    \caption{\textbf{(a)}: A $1$-dimensional potential $V(x)$ (inspired by the potential used in Refs.~\cite{Andreassen:2016cvx,Andreassen:2016cff}) with a barrier separating a false vacuum $\mathcal{F}$ from a true vacuum $\mathcal{R}$. For an example initial wavefunction $\psi(t=0)$, time evolution causes it to tunnel through the barrier, where the mass is fixed to $m = 6\cdot10^3$ in natural units ($\hbar = 1$). \textbf{(b)}: The time evolution of the probability $P_\mathcal{F}(t)$ of finding the particle in $\mathcal{F}$ is well-described by a 2-state resonant expansion, up to corrections at early times due to contributions from higher energy resonant states. At large times, the decay of $P_\mathcal{F}(t)$ approaches an exponential decay described by a single resonant state, shown in the inset figure. }
    \label{fig:onlyfig}
\end{figure*}

For simplicity, we focus in this work on the simple example of a point particle in one spatial dimension. However, due to our analysis' connection to the direct approach it inherits its straightforward generalisation to higher-dimensional systems, on which we comment in the discussion section. We denote the basin surrounding the false vacuum $x_{\rm FV}$ as $\mathcal{F}$, and its complement (the real vacuum region) by $\mathcal{R}$ (see \Cref{fig:onlyfig}{(a)}). We restrict ourselves to a brief but comprehensive discussion of the most important relations necessary to capture the early transient behavior, and provide extensive, detailed derivations in our companion article~\cite{Companion}.

{\it \bf Resonant Expansion.} 
The resonant expansion assumes that states $|\Psi\rangle$ localised within $\mathcal{F}$ can be expanded as a sum over normalised resonant states $|\tilde{E}_i\rangle$ \cite{Kobzarev:1974cp,Brezin:1976vw,Callan:1977pt,Zinn-Justin:2002ecy,TunnelingBook,Andreassen:2016cvx}, where $\tilde{E}_i = E_i + i\Gamma_i/2$ is the associated complex energy:
\begin{align}
    |\Psi\rangle= \sum_{i=1}^\infty \psi_i | \tilde{E}_i \rangle \, .
\end{align}
These states are constructed by imposing outgoing boundary conditions at the barrier, and because of that they are only approximate solutions to Schr\"odinger's equation --- of which exact solutions will feature some back-tunneling~\cite{Gamow:1928zz,Siegert:1939zz}. Ignoring the effects of back-tunneling (equivalently, requiring that $\Gamma_i > 0$ for all $i$), the resonant states are ordered such that $i < j$ implies $\Gamma_i < \Gamma_j$. In the long-time limit, the wavefunction is dominated by the lowest resonance $|\tilde{E}_1\rangle$, and the probability of finding the particle in the false vacuum $P_\mathcal{F}(t)$ decays exponentially in time:
\begin{equation}
P_\mathcal{F}(t) := \int_\mathcal{F} \mathrm{d}x\ |\psi(t)|^2, \quad  P_\mathcal{F}(t) \approx_{t \to \infty}  |\psi_1|^2  e^{- \Gamma_1 t}
\end{equation}
However, at pre-asymptotic times there are corrections due to the higher resonant states, which in general are non-orthogonal because of the non-hermiticity of the Hamiltonian restricted to the false-vacuum sector with outgoing boundary conditions. 

Including the effects of $n$ of the lowest resonant states, $P_\mathcal{F}(t)$ is given by the resonant expansion:
\begin{equation}\label{eq:fitform}
P_{\mathcal{F}}(t) \hspace{-0.1cm} \approx \hspace{-0.1cm} \sum_{i,j=1}^{n} {\rm Re} \left\{ \psi_{j}^* \langle \tilde{E}_j | \tilde{E}_i \rangle \psi_{i} \, e^{-i(E_i - E_j) t} \right\} e^{-(\Gamma_i + \Gamma_j) t/2} \, ,
\end{equation}
where due to the non-orthogonality of the resonant states, the non-diagonal overlap matrix $\langle \tilde{E}_j | \tilde{E}_i \rangle$ introduces \textit{oscillations} into the decay of $P_\mathcal{F}(t)$. This behaviour can be clearly seen in \Cref{fig:onlyfig}(b), where we show the probability contained within $\mathcal{F}$ as a function of time for the initial condition and potential shown in \Cref{fig:onlyfig}(a): following a short period in which there is contamination from higher-energy resonant states, the early-time decline of $P_{\mathcal{F}}$ is dominated by the contribution from the second resonance $|\tilde{E}_2\rangle$. As $\Gamma_2 >\Gamma_1$, this contribution becomes subdominant at larger times, as the second resonance has fully decayed. Between these two regimes we observe a transient dominated by the term mixing $\tilde{E}_1$ and $\tilde{E}_2$, manifesting in oscillatory behavior.

In principle, one could treat every variable in the resonant expansion Eq.~\eqref{eq:fitform} as a quantity to be predicted. However, the fact that the expansion takes this form is already predictive, as it implies a precise form of the time dependence that can be tested against numerical simulations, and also serves as a test of independent calculations of the resonant energies $E_i$ and tunneling rates $\Gamma_i$ --- e.g., those of the WKB approximation. The matrix elements $\langle \tilde{E}_i | \tilde{E}_j \rangle$ are also parameters characterising the false vacuum, and crucially account for the presence of the oscillatory behaviour in the probability. On the other hand, $\psi_{i,j}$ characterise the initial state. In practice, numerically fitting the resonant expansion as shown in \Cref{fig:onlyfig} works well, and the extracted values of ${E}_i$, $\Gamma_i$, and $\langle \tilde{E}_i | \tilde{E}_j \rangle$ are stable with respect to changing the initial state, as explored in our companion article~\cite{Companion}. This demonstrates that they are intrinsic properties of the system and independent of the initial condition, which one may seek to characterize with other methods, such as the \textit{steadyon} picture, whose applicability to this situation we establish in what follows. 

\begin{figure}[t]
\centering
\begin{tikzpicture}[scale=6]
  \draw[->] (0,0) -- (1.15,0) node[right] {$x$};
  \draw[->] (0,0) -- (0,0.4);
  \draw (0.81,0.41) node [black]  {$V(x)$}; 
\draw[domain=0:1.079,smooth,variable=\x,very thick] plot ({\x},{0.8*(\x)^2 - 0.6*(\x)^6});  
\draw[gray!60,blue,snake it, thick] (0,0.2)--(0.51,0.2);
  \draw[gray!60,dashed] (0.51,0)--(0.51,0.2);
  \draw[gray!60,dashed] (1,0)--(1,0.2);
  \draw (-0.01,-0.05) node [black] {$x_{\rm FV}$};
  \draw (0.51,-0.05) node [black]  {$x_*$};
  \draw (1,-0.05) node [black]  {$x_s$};
  \draw (0.3,0.4) node [black]  {$\mathcal{F}$};
  \draw (1.1,0.4) node [black]  {$\mathcal{R}$};
  \node at (0.51,0.2) [circle,draw=gray,fill=gray!60]{};
\end{tikzpicture}
\caption{At $t=0$, the particle is fully localised within the well $\mathcal{F}$. $x_*$ denotes the point beyond which its wave function decays exponentially. }
\label{fig:FRBarrier}
\end{figure}
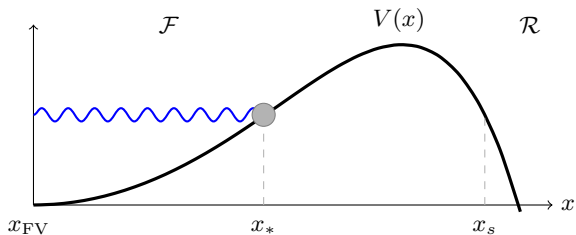

{\it \bf Probability flux. }
Our first step is to find an expression for the probability flux out of $\mathcal{F}$. For any state initialized within $\mathcal{F}$ and then evolved for a physical time $t$, we can define 
\begin{gather}
	P_{\mathcal{F}}(t) = P_{\mathcal{F}}(0) \cdot e^{- \Gamma(t) t} \, , \label{eq:GammaAnsatz} \\ 
    \Rightarrow \Gamma(t) = - \frac{\dot{P}_{\mathcal{F}} (t)}{P_{\mathcal{F}} (t)}= \frac{\dot{P}_{\mathcal{R}} (t)}{P_{\mathcal{F}} (t)} \, ,\label{eq:GammaDef}
\end{gather}
where in the last step we have used that the probability flow out of $\mathcal{F}$ is the same as that into $\mathcal{R}$. In an exact resonance state $|\tilde{E}\rangle$ with $\tilde{E}=E+ i\Gamma_i/2$, we have $\Gamma(t) = \Gamma_i$. These probabilities can be expressed in terms of the theory's Feynman propagator $D_F$ as
\begin{align}
    &P_{\Omega} (t) = \hspace{-0.1cm} \int_{\Omega} \hspace{-0.1cm}  \d x\ \langle x | e^{-i H t} | \Psi \rangle \langle \Psi | e^{i H t}|x\rangle \nonumber \\ 
    &= \hspace{-0.1cm}  \int_{\Omega}\hspace{-0.1cm}  \d x\hspace{-0.1cm}  \int \hspace{-0.1cm}  \d x_1 \hspace{-0.1cm}  \int \hspace{-0.1cm}  \d x_2\ \psi(x_1) \psi^* (x_2) \langle x | e^{-i H t} | x_1 \rangle \langle x_2 | e^{i H t}|x\rangle \nonumber \\
    &=\hspace{-0.1cm}  \int_{\Omega} \hspace{-0.1cm}  \d x \hspace{-0.1cm}  \int \hspace{-0.1cm}  \d x_1 \hspace{-0.1cm}  \int \hspace{-0.1cm}  \d x_2\ \psi(x_1) \psi^* (x_2) D_F (x,t|x_1) D_F^* (x,t|x_2) \label{eq:Pgen} \, ,
\end{align}
whose structure we will use to calculate $\dot{P}_{\mathcal{R}} (t)$ and $P_{\mathcal{F}} (t)$. Let us consider the case where $\psi$ is exactly the wavefunction of a resonance state. First, in order to calculate $\dot{P}_{\mathcal{R}} (t)$, we follow Ref.~\cite{Andreassen:2016cff} and introduce the auxiliary propagator $\bar{D}_F$ defined by
\begin{align}
    \bar{D}_F (x_s , t_s | x_i) \equiv 
    \sint{x (0) = x_i}{x (t_s) = x_s}{1} {\mathcal D} x \ e^{i S[x]} \delta \left( F_{x_s}[x] - t_s \right) \, , 
\end{align}
which satisfies
\begin{align}
    D_F(x_f,t | x_i)= \int_{0}^t \d t_s D_F( x_f, t |x_s, t_s ) \ \bar{D}_F(x_s,t_s|x_i) \,.
\end{align}
The point $x_s$ is determined by the energy $E$ of the resonance state, as we now explain by means of the illustration in Fig.~\ref{fig:FRBarrier}. Within the basin $\mathcal{F}$, the wave function $\psi$ of the resonance state is localised within a region confined on one side by $x_*$, defined as the solution to 
$V(x_*)=E$ on the $\mathcal{F}$ side of the barrier. We define
$x_s$ as the point that is energetically degenerate with $x_*$ on the other side of the barrier. The functional $F_{x_s}[x(t)]$ then maps the path $x(t)$ onto the time when it first crosses the point $x_s$ --- i.e., when the analog classical particle completely crosses the barrier. 

Next, we make use of the first implicit assumption underlying Eq.~\eqref{eq:GammaAnsatz}, namely the absence of back-tunneling. In terms of the propagator, this assumption translates to asserting that the propagation of particles from $x_s$ back into $\mathcal{F}$ is negligible, and thus the integration over $\mathcal{R}$ in $\dot{P}_\mathcal{R}$ may be extended to all of space
\begin{align}
    \int_\mathcal{R}\d x_f\ &  D_F (  x_f , t | x_s , t_s )  D_F^* (  x_f , t | x_{s} , t_s^\prime )  \nonumber \\ 
    &\simeq  \int \d x_f\  D_F (  x_f , t | x_s , t_s )  D_F^* (  x_f , t | x_{s} , t_s^\prime ) \label{eq:nobacktunneling} \\
    &=  D_F ( x_s , t_s^\prime | x_s , t_s  ) \, . \nonumber
\end{align}
It is then straightforward to show that 
\begin{align}
	\dot{P}_{\mathcal{R}} (t)& = \int \d x_1 \int \d x_2 \ \psi(x_1) \psi^* (x_2) \label{eq:PRdotint} \\ 
     &\times  D_F (  x_s , t | x_1 ) \bar{D}_F^* ( x_s ,t | x_2 ) +c.c. \nonumber 
\end{align}
To further develop this expression we introduce two more auxiliary propagators,
\begin{align}
    \underaccent{\bar}{D}_F ( x_s,t_s | x_* ,t_*) \equiv & \quad \sint{x(t_*) = x_*}{x(t_s) = x_s}{1} \mathcal{D}x \ e^{iS[x]} \delta (G_{x_*} [x] -t_*) \, , \\ 
    \bar{\underaccent{\bar}{D}}_F ( x_s,t_s | x_* ,t_* )  \equiv &\\ 
     \quad  \sint{x(t_*) = x_*}{x(t_s) = x_s}{1} \mathcal{D}x \ e^{iS[x]} & \delta (G_{x_*} [x] -t_*)  \delta (F_{x_s} [x] -t_s)  \, .\nonumber
\end{align}
The functional $G_{x_*}[x]$ now maps the path $x(t)$ onto the time the path $x(t)$ crosses $x_*$ for the \textit{last} time. These new functions can be related to $D_F$ and $\bar{D}_F$ through the integral relations
\begin{align}
    D_F (  x_s , t | x_i )= & \int_0^t \d t_* \  \underaccent{\bar}{D}_F (x_s,t | x_* ,t_*) D_F(x_*,t_* | x_i) \, , \\
    \bar{D}_F ( x_s , t | x_i )= & \int_0^t \d t_* \  \bar{\underaccent{\bar}{D}}_F ( x_s,t | x_* ,t_*) D_F(x_*,t_* | x_i) \, .
\end{align}
These decompositions require that $x_i < x_* <x_s$, which can strictly only be achieved if these expressions were to be integrated against wave functions that are fully localised within $\mathcal{F}$. For a resonance state $|\tilde{E}\rangle$, $x_*$ is the classical turning point of a classical particle with energy $E$, beyond which the state's wave function decays exponentially. This makes transparent the implicit assumption of the tunneling process at hand --- i.e., that our treatment assumes that most of the probability is contained within $\mathcal{F}$ before reaching $x_*$. 

When applicable, these decompositions allows one to recombine the wave function with one of the emerging propagators through
\begin{align}\label{eq:propInt}
    \int \d x_1 D_F(x_*,t_*|x_1) \psi(x_1)  = \psi (x_*, t_*) \, .
\end{align}
This further simplifies Eq.~\eqref{eq:PRdotint} to
\begin{align}
    \dot{P}_{\mathcal{R}} = & \int_0^t \d \Delta t  \ \underaccent{\bar}{D}_F (x_s,\Delta t | x_* ) \psi (x_*,t-\Delta t) \label{eq:P-flux-master} \\ 
    \times & \int_0^t \d \Delta t^\prime \ \bar{\underaccent{\bar}{D}}_F^* (x_s, \Delta t^\prime | x_* ) \psi^* (x_*,t-\Delta t^\prime)  +c.c. \,. \nonumber 
\end{align}
While we obtained this relation with the motivation of calculating $\Gamma$ for a resonance state through Eq.~\eqref{eq:GammaDef}, it does not rely on this context. In fact, it holds true for \textit{any} wave function subject to our previously made assumptions --- the localisation within $\mathcal{F}$ beyond some point $x_* \in \mathcal{F}$ and the absence of back-tunneling. Moreover, it is \textit{exact} beyond these assumptions, not relying on any further expansion.

{\it \bf Semi-classical evaluation.} The propagators in Eq.~\eqref{eq:P-flux-master} can be evaluated semi-classically in the so-called \textit{steadyon picture} developed in Refs.~\cite{Steingasser:2024ikl,Steingasser:2023gde}. These works suggest regularizing the theory by a deformation of the Hamiltonian by introducing an infinitesimal imaginary part, $H\to (1-i \epsilon)H$. In this deformed theory, the path integrals representing the propagators can be evaluated through a stationary phase approximation around a complex solution $\bar{x}(t)$,
\begin{align}
    \underaccent{\bar}{D}_F (x_s,\Delta t | x_* ) \sim \bar{\underaccent{\bar}{D}}_F^* (x_s, \Delta t | x_* ) \sim e^{i S[\bar{x}]} \,.
\end{align}
The complex solution $\bar{x}(t)=\bar{x}_{\rm Re}(t) + \bar{x}_{\rm Im}(t)$, called \textit{steadyon}, emerges as a solution to the complexified equation of motion
\begin{align}\label{eq:eomclass}
    \ddot{\bar{x}}(t^\prime)=(1-2 i \epsilon )V^\prime (\bar{x}(t^\prime))\,.
\end{align}
This equation alone determines $\bar{x}(t)$ only up to its boundary conditions, which naturally depend on the time interval $\Delta t$. Thus, the integral over $\Delta t$ in Eq.~\eqref{eq:P-flux-master} amounts to an integral over contributions from different steadyons. In addition to the direct dependence of the action on $\Delta t$, this induces an indirect dependence of the action on $\Delta t$ through the form of the steadyon.

To identify the dominant contribution, we can evaluate this integral by first carrying out another saddle point (stationary phase) approximation for the combination of propagator with wave function. For the latter, we find (with the $\epsilon$-regularisation of the Hamiltonian)
\begin{align}
    \psi (x_*, t-\Delta t) \simeq e^{-i E_i t - \epsilon E_i t } \cdot e^{i E_i \Delta t  + \epsilon E_i \Delta t } \cdot  \psi(x_*) \, ,
\end{align}
which is nothing more than the usual time evolution for a state with energy $E_i$ in the $\epsilon$-deformed theory, neglecting the (small) effect of the (yet-to-be-calculated) tunneling rate $\Gamma_i$.\footnote{See our companion paper~\cite{Companion} for a detailed discussion of this point.}
In order to make manifest the complex nature of the steadyon's action, as well as its dependence on $\Delta t$, we introduce the decomposition $S[\bar{x}]=S_{\rm Re} (\Delta t)+i S_{\rm Im} (\Delta t)$. It then follows that the integrals in Eq.~\eqref{eq:P-flux-master} are extremised by steadyons satisfying the equations
\begin{gather}\label{eq:extconds1}
    \frac{\d}{\d \Delta t} S_{\rm Re} (\Delta t) = - E_i \, , \\
    \quad \frac{\d}{\d \Delta t} S_{\rm Im} (\Delta t) = \epsilon  E_i\, . \label{eq:extconds2}
\end{gather}
In the following, we will argue that these conditions are satisfied by configurations dubbed \textit{periodic steadyons} in Ref.~\cite{Steingasser:2024ikl}. These are solutions to Eq.~\eqref{eq:eomclass} characterized by the additional boundary conditions $\dot{\bar{x}}(0)=\dot{\bar{x}}(\Delta t_{\rm per})=0$. These solutions only satisfy the correct boundary conditions for a suitable time $\Delta t_{\rm per}$ which depends on $\epsilon$ through $\Delta t_{\rm per} = N \epsilon^{-1} t_{\rm sys}$, where $N$ is an $\mathcal{O}(1)$ coefficient and $t_{\rm sys}$ the system's typical time scale, in agreement with our previous interpretation of the integral over $\Delta t$ as an integral over different saddle points.

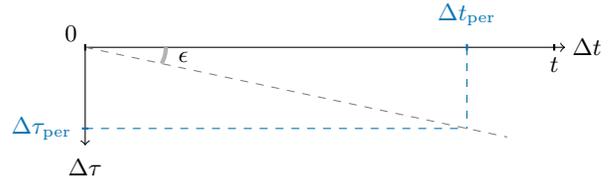
\begin{figure}[t!]
\begin{minipage}[c]{0.52\textwidth}
\centering
\vspace*{-0.5cm}
\hspace*{-0.7cm}
\begin{tikzpicture}[scale=1.45]
    \draw[->](0,0)--(4.4,0); 
    \draw[->](0,0)--(0,-0.9); 
    \draw[gray, dashed](0,0)--(3.87, -0.82341); 
    \draw[black, thick](0,0.03)--(0,-0.03);
    \draw (-0.13,0.13) node [black]{$0$};
    \draw (4.3,-0.16) node [black]{$t$};
    \draw (4.6,0) node [black]{$\Delta t$};
    \draw[black, thick](4.3,0.03)--(4.3,-0.03);
    \draw (0,-1.1) node [black]{$\Delta \tau$};
    \draw (0.9,-0.09) node [black] {$\epsilon$};
    \draw [gray!60,ultra thick](0.74,0) arc [start angle=0, end angle=-30, radius=0.3];
    \draw[oiBlue, dashed](3.5,0)--(3.5,-0.7447); 
    \draw[oiBlue, dashed](0,-0.7447)--(3.5,-0.7447);
    \draw[oiBlue, thick](3.5,0.03)--(3.5,-0.03);
    \draw[oiBlue, thick](-0.03,-0.7447)--(0.03,-0.7447);
    \draw (-0.4,-0.7447) node [text=oiBlue]{$\Delta \tau_{\rm per}$}; 
    \draw (3.5,0.3) node [text=oiBlue]{$\Delta t_{\rm per} $}; 
\end{tikzpicture}
\end{minipage}
\caption{Assigning the factor $1-i \epsilon$ to the time variable rather than the Hamiltonian allows to interpret the steadyon as a solution along a diagonal contour in the complex time plane. Its projections on the real- and imaginary-time axis, respectively, yield the familiar classical and Euclidean-time solutions, respectively.}
\label{fig:Wick}
\end{figure}

We first observe that the deformation of the Hamiltonian is equivalent to an infinitesimal Wick rotation for systems without an explicit time-dependence. In this picture the steadyon emerges as a solution of its respective equation on a rotated contour in the complex-time plane, and its action as a line integral over this contour. As both $\bar{x}(t)$ and its action are analytic functions, its action is invariant under a deformation of this contour into one segment parallel to the real-time axis and one parallel to the Euclidean-time axis. See Fig.~\ref{fig:Wick}. We will now show that the contribution from the periodic steadyon's projection onto the real-time segment exactly matches Eq.~\eqref{eq:extconds1}, while its Euclidean-time counterpart yields Eq.~\eqref{eq:extconds2}.

Using the periodic steadyon's characteristic boundary condition $\dot{\bar{x}}(0)=0$ together with $\epsilon=0$ (corresponding to the contour in real-time direction), it is easy to see that Eq.~\eqref{eq:eomclass} reduces to the classical equation of motion for the real part of the steadyon. Thus, its projection onto this contour can be identified with the classical, oscillatory motion with turning point $x_*$, $x_{\rm class}(t)$. If $\Delta t=\Delta t_{\rm per}$, this solution by construction satisfies the necessary boundary conditions. Recall now that the action $S_{\rm Re}(\Delta t)$ depends on $\Delta t$ both directly through the time integral and indirectly through the change in $\bar{x}$ necessary to accommodate the boundary conditions. For $\Delta t= \Delta t_{\rm per} - \delta t$, it is then straightforward to show that the action changes as $S_{\rm Re}\to S_{\rm Re}+ \delta t \cdot V(x_*)$, reproducing Eq.~\eqref{eq:extconds1}. Moreover, the identification of the periodic steadyon along this contour with the classical solution implies that its contribution to the Euclidean action is strictly real.

Along the segment parallel to the Euclidean-time axis, Eq.~\eqref{eq:eomclass} reduces to the Euclidean equation of motion, and the action calculated along it to the Euclidean action up to a factor of $i$ from the line element. The condition $\dot{\bar{x}}(0)$ meanwhile again implies a motion out of rest. Taken together, these simple observations imply that the projection of the steadyon onto the Euclidean-time segment reproduces the periodic Euclidean-time instanton. Its contribution to the overall action is therefore the Euclidean action $S_E$ of this solution times a factor of $i$. This, in turn, implies that $S_{\rm Im}=S_E$. As the length of the Euclidean-time interval is related to its real-time counterpart through $\d \Delta \tau = - \epsilon \d \Delta t$, it is easy to see that $S_{\rm Im}$ satisfies Eq.~\eqref{eq:extconds2} for the same reasons as $S_{\rm Re}$ satisfies Eq.~\eqref{eq:extconds1}. 

Having identified the dominant contribution to the $\Delta t$-integrals, it is straightforward to obtain $\dot{P}_{\mathcal{R}}$ to leading order as 
\begin{align}\label{eq:PRdotstat}
    \dot{P}_{\mathcal{R}} \sim & e^{-2 \epsilon E_i t} \exp \left( - S_E [x_{b,\rm{per}}]  + 2  E_i \Delta \tau_{\rm per} \right)\, ,
\end{align}
where $x_{b,\rm{per}}$ denotes the \textit{periodic bounce solution} describing the motion from $x_*$ to $x_s$ and back. 

To obtain the tunneling rate through Eq.~\eqref{eq:PRdotint}, it remains to calculate the probability $P_{\mathcal{F}}$ taking into account the deformation of the Hamiltonian. Using Eq.~\eqref{eq:PRdotstat}, we find:
\begin{align}
    P_\mathcal{F} (t)=& \int_\mathcal{F} |\psi_i (t,x)|^2 \d x \nonumber \\ 
    =& \int_{\mathcal{F}} e^{- 2 \epsilon E_i t} |\psi_i (x)|^2 \simeq e^{- 2 \epsilon E_i t} \, .
\end{align}
In the last step, we have made use of the approximation that a majority of the wave function remains in the false vacuum basin $\mathcal{F}$. Thus, we find that this approach is to leading order indeed capable of recovering the familiar result obtained by relying on the WKB approximation,
\begin{align}\label{eq:WKB}
    \Gamma  =  \frac{\dot{P}_{\mathcal{R}}(t)}{P_{\mathcal{F}}(t)} \simeq  |\psi_i(x_*)|^2 \exp \left( - S_E [x_{b,\rm{per}}]  + 2  E_i \Delta \tau_{\rm per} \right) \,.
\end{align}
{\it \bf Discussion.} We have shown that tunneling at pre-asymptotic times can be described in terms of the tunneling rates out of linear combinations of individual resonance states. With the decomposition in terms of these states being equivalent to specifying the initial state, a complete analysis of tunneling during these times thus amounts to the calculation of these rates. We have provided a first-principles derivation of these rates, combining the direct approach of~\cite{Andreassen:2016cvx,Andreassen:2016cff} with the more recently developed steadyon picture~\cite{Steingasser:2024ikl,Steingasser:2023gde}.

Our analysis inherits and expands the advantages of the direct approach analysis of the asymptotic false vacuum rate. 

Unlike the analysis based on the WKB approximation, it makes evident the role of physical time for the tunneling. On a conceptual level, it establishes a clear connection between real and imaginary time, generalizing the Wick-rotation based argument for the false vacuum analysis. Moreover, it allows for a clear interpretation of the relevant time scales. The ability of the steadyons to serve as approximate saddle points of the real-time path integrals that they are used to evaluate requires $t\gg t_{\rm sys}$. The existence of an upper bound on the time $t$ where this approximation is valid follows from two factors. First, we demanded no back-tunneling, which was already responsible for this bound for the false vacuum analysis. However, our analysis illustrates a second factor: The periodic instanton emerges only after specifying that the initial state is an approximate eigenstate of the Hamiltonian. As soon as the resonance ``disintegrates'' the factor $\psi(x_*, \Delta t)$ changes, and the periodic instanton is no longer the optimal solution. Having this clear form, however, in principle also allows one to interpolate beyond the disintegration of the resonance state, given knowledge of how $\psi(x_*, \Delta t)$ evolves past this point. 

More generally, our approach is easily generalized to more general states, for which the WKB approximation might not work, enabling a direct calculation without the need to first decompose the state of interest in terms of resonances.

The clarity of assumptions underlying our analysis makes it well-suited for precision calculations. Even though we restricted ourselves to the calculation of the leading-order exponent, our expression for the probability flux in Eq.~\eqref{eq:P-flux-master} is true non-perturbatively. It thus naturally allows for a systematically improvable perturbative treatment. The representation of our results in terms of path integrals also simplifies investigations concerning gauge invariance~\cite{Metaxas:1995ab,Andreassen:2014gha,Andreassen:2014eha}, as well as calculations in situations in which resummations are necessary to address large back-reactions~\cite{Coleman:1973jx}, e.g., for finite-temperature processes~\cite{Steingasser:2023gde}. 

Our analysis also inherits the direct approach's simple generalisation to higher-dimensional systems by integrating the points $x_*$ and $x_s$ over the submanifolds of energetically degenerate points, compared to the more involved procedure necessary for a direct application of the WKB method in multiple dimensions~\cite{Banks:1973ps}. The resulting expression can be understood as the overall tunneling rate being described by a collection of instantons from all classical turning points to any of the energetically degenerate points beyond the barrier. Our analysis moreover makes explicit the definition of the regions $\mathcal{F}$ and $\mathcal{R}$ through the localisation of the wave function.

\section*{Acknowledgements}

We acknowledge the hospitality of the Munger residence and the Kavli Institute for Theoretical Physics at the University of California, Santa Barbara, where parts of this research were conducted. We furthermore thank Matt D. Charles for significant contributions to an inspiring work environment and Sebastian Schenk for helpful discussions.

Portions of this work were conducted in MIT's \textit{Center for Theoretical Physics - a Leinweber institute} and partially supported by the U.S.~Department of Energy under Contract No.~DE-SC0012567. 
While at MIT, JL's contributions were supported by the U.S. Department of Energy under Contract No.~DE-SC0011090, by the SciDAC5 award DE-SC0023116, and additionally by the National Science Foundation under Cooperative Agreement PHY-2019786 (The NSF AI Institute for Artificial Intelligence and Fundamental Interactions, http://iaifi.org/). While at Argonne National Laboratory, JL's contributions were supported by the U.S. Department of Energy, Office of Science, Office of Nuclear Physics through Contract No.~DE-AC02-06CH11357. 

This research was supported in part by grant NSF PHY-2309135 to the Kavli Institute for Theoretical Physics (KITP). BSH's contributions were also supported by grant 994312 from the Simons Foundation.

TS's contributions to this work were made possible by the Walter Benjamin Programme of the Deutsche Forschungsgemeinschaft (DFG, German Research Foundation)
- 512630918. This project was also supported in part by the Black Hole Initiative at Harvard University, with support from the Gordon and Betty Moore Foundation and the John Templeton Foundation. 
He also acknowledges partial financial support from the Spanish Research Agency (Agencia Estatal de Investigaci\'on) through the grant IFT Centro de Excelencia Severo Ochoa No CEX2020-001007-S and PID2022-137127NB-I00 funded by MCIN/AEI/10.13039/501100011033/ FEDER, UE. This project has received funding/support from the European Union's Horizon 2020 research and innovation programme under the Marie Sklodowska-Curie Staff Exchange  grant agreement No.~101086085 -ASYMMETRY.
The opinions expressed in this publication are those of the authors and do not necessarily reflect the views of these Foundations.

\bibliography{main}

\end{document}